\documentclass[12pt,english,journal,12pt,onecolumn,draftclsnofoot]{IEEEtran}
\usepackage[T1]{fontenc}
\usepackage[latin9]{inputenc}
\usepackage{geometry}
\usepackage{color}
\usepackage{amstext}
\usepackage{graphicx}
\usepackage{setspace}
\usepackage{esint}
\makeatletter
\@ifundefined{showcaptionsetup}{}{%
 \PassOptionsToPackage{caption=false}{subfig}}
\usepackage{subfig}
\makeatother

\usepackage{babel}
\begin{document}
\title{A Fast, Accurate, and Separable Method for Fitting a Gaussian Function}
\author{Ibrahim Al-Nahhal, Octavia A. Dobre, Ertugrul Basar, Cecilia Moloney,
and Salama Ikki}

\maketitle
\vspace{-10mm}

The Gaussian function (GF) is widely used to explain the behavior
or statistical distribution of many natural phenomena as well as industrial
processes in different disciplines of engineering and applied science.
For example, the GF can be used to model an approximation of the Airy
disk in image processing, laser heat source in laser transmission
welding \cite{key-1}, practical microscopic applications \cite{key-2},
and fluorescence dispersion in flow cytometric DNA histograms \cite{key-3}.
In applied sciences, the noise that corrupts the signal can be modeled
by the Gaussian distribution according to the central limit theorem.
Thus, by fitting the GF, the corresponding process/phenomena behavior
can be well interpreted.

This article introduces a novel fast, accurate, and separable algorithm
for estimating the GF parameters to fit observed data points. A simple
mathematical trick can be used to calculate the area under the GF
in two different ways. Then, by equating these two areas, the GF parameters
can be easily obtained from the observed data.

\vspace{-2mm}

\section*{Gaussian Function Fitting Approaches}

A GF has a symmetrical bell-shape around its center, with a width
that smoothly decreases as it moves away from its center on the $x$-axis.
The mathematical form of the GF is

\vspace{-5mm}

\begin{equation}
y(x)=Ae^{-\frac{(x-\mu)^{2}}{2\sigma^{2}}},\label{eq:1 gauss form}
\end{equation}

\vspace{-1mm}

\noindent with three shape-controlling parameters, $A$, $\mu$ and
$\sigma$, where $A$ is the maximum height (amplitude) that can be
achieved on the $y$-axis, $\mu$ is the curve-center (mean) on the
$x$-axis, and $\sigma$ is the standard deviation (SD) which controls
the width of the curve along the $x$-axis. The aim of this article
is to present a new method for the accurate estimation of these three
parameters. The difficulty of this lies in estimating the three shape-controlling
parameters ($A$, $\mu$ and $\sigma$) from observations, that are
generally noisy, by solving an over-determined nonlinear system of
equations.

The standard solutions for fitting the GF parameters from noisy observed
data are obtained by one of the following two approaches:

\vspace{-1mm}

\begin{enumerate}
\item Solving the problem as a nonlinear system of equations using one of
the least-squares optimization algorithms. This solution employs an
iterative procedure such as the Newton-Raphson algorithm \cite{key-4}.
The drawbacks of this approach are the iterative procedure, which
may not converge to the true solution, as well as its high cost from
the computational complexity perspective.
\item Solving the problem as a linear system of equations based on the fact
that the GF is an exponential of a quadratic function. By taking the
natural logarithm of the observed data, the problem can be solved
in polynomial time as a $3\times3$ linear system of equations. Two
traditional algorithms have been proposed in this context: Caruana's
algorithm \cite{key-5} and Guo's algorithm \cite{key-6}. Furthermore,
instead of taking the natural logarithm, the partial derivative is
used in Roonizi's algorithm \cite{key-7}.
\end{enumerate}
In this article, we will consider only the second approach, which
is more suitable for most scientific applications, due to its simplicity
and avoidance of the drawbacks of the first approach. Let us start
with a brief introduction of the existing three algorithms for the
second approach.

\vspace{-2mm}

\section*{Caruana's Algorithm}

Caruana's algorithm exploits the fact that the GF is an exponential
of a quadratic function and transforms it into a linear form by taking
the natural logarithm of (\ref{eq:1 gauss form}) to obtain

\vspace{-5mm}

\begin{equation}
\text{ln}(y)=\text{ln}(A)+\frac{-(x-\mu)^{2}}{2\sigma^{2}}=\text{ln}(A)-\frac{\mu^{2}}{2\sigma^{2}}+\frac{2\mu x}{2\sigma^{2}}-\frac{x^{2}}{2\sigma^{2}}=a+bx+cx^{2},\label{eq: linear guass}
\end{equation}

\noindent where $a=\text{ln}(A)-\mu^{2}/\left(2\sigma^{2}\right)$,
$b=\mu/\sigma^{2}$ and $c=-1/\left(2\sigma^{2}\right)$. Accordingly,
the unknowns become $a$, $b$ and $c$ in the linear equation (\ref{eq: linear guass})
instead of $A$, $\mu$ and $\sigma$ in the nonlinear equation (\ref{eq:1 gauss form}).
Next, if the observations $y$ are noisy, then they can be modeled
as $\hat{y}=y+w$; each contains the ideal data point, $y$, that
is corrupted by the noise, $w$ with SD of $\sigma_{w}$. Note that
in (\ref{eq: linear guass}), we consider only the observations that
have values above zero.

Once we have an over-determined linear system, the unknowns can be
estimated using the least-squares method. Caruana's algorithm estimates
the three unknowns ($a$, $b$ and $c$) in (\ref{eq: linear guass})
using the least-squares method by forming the error function, $\varepsilon$,
for (\ref{eq: linear guass}) as

\vspace{-4mm}

\begin{equation}
\varepsilon=\text{ln}(\hat{y})-\text{ln}(y)=\text{ln}(\hat{y})-(a+bx+cx^{2}).\label{eq:3 error}
\end{equation}

\vspace{-2mm}

\noindent Then, by differentiating the sum of $\varepsilon^{2}$ with
respect to $a$, $b$ and $c$ and equating the results to zero, we
obtain three equations, which represent the following linear system

\vspace{-2mm}

\begin{equation}
\left[\begin{array}{ccc}
N & \sum x_{n} & \sum x_{n}^{2}\\
\sum x_{n} & \sum x_{n}^{2} & \sum x_{n}^{3}\\
\sum x_{n}^{2} & \sum x_{n}^{3} & \sum x_{n}^{4}
\end{array}\right]\left[\begin{array}{c}
a\\
b\\
c
\end{array}\right]=\left[\begin{array}{c}
\sum\text{ln}(\hat{y}_{n})\\
\sum x_{n}\text{ln}(\hat{y}_{n})\\
\sum x_{n}^{2}\text{ln}(\hat{y}_{n})
\end{array}\right],\label{eq:7 car matrix}
\end{equation}

\noindent where $N$ is the number of observed data points and $\sum$
denotes $\sum_{n=1}^{N}$. In this case, the parameters $a$, $b$
and $c$ can be determined simply by solving (\ref{eq:7 car matrix})
as a determined linear system of equations. Subsequently, the original
parameters of the GF are determined as

\vspace{-3mm}

\begin{singlespace}
\begin{equation}
A=e^{a-\frac{b^{2}}{4c}},\,\,\,\,\,\,\,\mu=\frac{-b}{2c},\,\,\,\,\,\,\,\sigma=\sqrt{\frac{-1}{2c}}.\label{eq:8 mu =000026 sigma =000026 A}
\end{equation}

\end{singlespace}

The weighted least-squares method is the second candidate method to
estimate the unknowns, and it is expected to have a better accuracy
in estimation rather than the least-squares method.

\vspace{-2mm}

\section*{Guo's Algorithm}

Guo's algorithm is a modified version of the Caruana\textquoteright s
algorithm, which finds the unknowns $a$, $b$ and $c$ in (\ref{eq: linear guass})
using the weighted least-squares method. It uses the noisy observed
data, $\hat{y}$, to weight the error function in (\ref{eq:3 error}).
Therefore, the error equation in (\ref{eq:3 error}) becomes $\delta=\hat{y}\varepsilon=\hat{y}[\text{ln}(\hat{y})-(a+bx+cx^{2})]$,
and the linear system of equations in (\ref{eq:7 car matrix}) becomes

\vspace{-3mm}

\begin{equation}
\left[\begin{array}{ccc}
\sum\hat{y}_{n}^{2} & \sum x_{n}\hat{y}_{n}^{2} & \sum x_{n}^{2}\hat{y}_{n}^{2}\\
\sum x_{n}\hat{y}_{n}^{2} & \sum x_{n}^{2}\hat{y}_{n}^{2} & \sum x_{n}^{3}\hat{y}_{n}^{2}\\
\sum x_{n}^{2}\hat{y}_{n}^{2} & \sum x_{n}^{3}\hat{y}_{n}^{2} & \sum x_{n}^{4}\hat{y}_{n}^{2}
\end{array}\right]\left[\begin{array}{c}
a\\
b\\
c
\end{array}\right]=\left[\begin{array}{c}
\sum\hat{y}_{n}^{2}\text{ln}(\hat{y}_{n})\\
\sum x_{n}\hat{y}_{n}^{2}\text{ln}(\hat{y}_{n})\\
\sum x_{n}^{2}\hat{y}_{n}^{2}\text{ln}(\hat{y}_{n})
\end{array}\right].\label{eq:16 matrix guo}
\end{equation}

\vspace{-1mm}

\noindent Moreover, the values of $A$, $\mu$ and $\sigma$ can be
computed from (\ref{eq:8 mu =000026 sigma =000026 A}).

One of the problems that affects the estimation accuracy is the \textit{long
tail} GF. This is experienced when the number of small values in the
observed data is large compared to the observed data length, $N$,
which means that a large noise exists in those observations. Thus,
an iterative procedure is required to improve the estimation accuracy.

\vspace{-2mm}

\section*{\label{subsec:Guo's-Algorithm-with iterative}Guo's Algorithm with
Iterative Procedure}

The estimation accuracy of the Guo's algorithm deteriorates for a
long tail GF. In order to increase the accuracy of fitting the long
tail Gaussian parameters, an iterative procedure for (\ref{eq:16 matrix guo})
is given as

\vspace{-4mm}

\begin{equation}
\left[\begin{array}{ccc}
\sum\hat{y}_{n,(k-1)}^{2} & \sum x_{n}\hat{y}_{n,(k-1)}^{2} & \sum x_{n}^{2}\hat{y}_{n,(k-1)}^{2}\\
\sum x_{n}\hat{y}_{n,(k-1)}^{2} & \sum x_{n}^{2}\hat{y}_{n,(k-1)}^{2} & \sum x_{n}^{3}\hat{y}_{n,(k-1)}^{2}\\
\sum x_{n}^{2}\hat{y}_{n,(k-1)}^{2} & \sum x_{n}^{3}\hat{y}_{n,(k-1)}^{2} & \sum x_{n}^{4}\hat{y}_{n,(k-1)}^{2}
\end{array}\right]\left[\begin{array}{c}
a_{(k)}\\
b_{(k)}\\
c_{(k)}
\end{array}\right]=\left[\begin{array}{c}
\sum\hat{y}_{n,(k-1)}^{2}\text{ln}(\hat{y}_{n})\\
\sum x_{n}\hat{y}_{n,(k-1)}^{2}\text{ln}(\hat{y}_{n})\\
\sum x_{n}^{2}\hat{y}_{n,(k-1)}^{2}\text{ln}(\hat{y}_{n})
\end{array}\right],\label{eq:iterative Guo matrix}
\end{equation}

\noindent where $\hat{y}_{n,(k)}=\hat{y}_{n}$ for $k=0$ and $\hat{y}_{n,(k)}=e^{a_{(k)}+b_{(k)}x_{n}+c_{(k)}x_{n}^{2}}$
for $k>0$, with the parenthesized subscripts denoting the indices
of iteration.

\vspace{-4mm}

\section*{Roonizi's Algorithm}

Roonizi's algorithm is designed to fit the GF riding on a polynomial
background. It can be utilized to fit a GF by taking the partial derivative
of (\ref{eq:1 gauss form}), and then taking the integral of the result
to obtain

\vspace{-2.5mm}

\begin{equation}
y(x)=\beta_{1}\phi_{1}(x)+\beta_{2}\phi_{2}(x),\label{eq:Roonizi_y}
\end{equation}

\vspace{-3.5mm}

\noindent where $\beta_{1}=-1/\sigma^{2}$, $\beta_{2}=\mu/\sigma^{2}$,
and

\vspace{-2.5mm}

\begin{equation}
\phi_{1}(x)=\int_{-\infty}^{x}u\,y(u)\,\,du,\,\,\,\,\,\,\,\,\phi_{2}(x)=\int_{-\infty}^{x}y(u)\,\,du.\label{eq: Phi_Roonizi}
\end{equation}

\vspace{-2.5mm}

\noindent Similar to the steps in Caruana's and Guo's algorithms,
to obtain the linear system of equations, the error of (\ref{eq:Roonizi_y})
becomes $\zeta=\hat{y}-(\beta_{1}\phi_{1}(x)+\beta_{2}\phi_{2}(x))$,
and a linear system of equations results as

\vspace{-2.5mm}

\begin{equation}
\left[\begin{array}{cc}
\sum\left|\phi_{1}(x_{n})\right|^{2} & \sum\phi_{1}(x_{n})\phi_{2}(x_{n})\\
\sum\phi_{1}(x_{n})\phi_{2}(x_{n}) & \sum\left|\phi_{2}(x_{n})\right|^{2}
\end{array}\right]\left[\begin{array}{c}
\beta_{1}\\
\beta_{2}
\end{array}\right]=\left[\begin{array}{c}
\sum\phi_{1}(x_{n})\hat{y}_{n}\\
\sum\phi_{2}(x_{n})\hat{y}_{n}
\end{array}\right].\label{eq: matrix_Roonizi}
\end{equation}

\vspace{-2.5mm}

\noindent By solving (\ref{eq: matrix_Roonizi}) in terms of $\beta_{1}$
and $\beta_{2}$, the estimated $\hat{\mu}$ and $\hat{\sigma}$ of
the GF can be calculated as

\vspace{-2.5mm}

\begin{equation}
\hat{\sigma}=\sqrt{\frac{-1}{\beta_{1}}},\,\,\,\,\,\,\,\,\,\hat{\mu}=\frac{-\beta_{2}}{\beta_{1}}.\label{eq: Mu_Sigma_Roonizi}
\end{equation}

\vspace{-2.5mm}

\noindent Finally, using $\hat{\mu}$ and $\hat{\sigma}$ from (\ref{eq: Mu_Sigma_Roonizi}),
the estimated $\hat{A}$ of the GF can be calculated as

\vspace{-2.5mm}

\begin{equation}
\hat{A}=\frac{\sum\left(\hat{y}_{n}\,\text{exp}\left(\frac{-(x_{n}-\hat{\mu})^{2}}{2\hat{\sigma}^{2}}\right)\right)}{\sum\text{exp}\left(\frac{-(x_{n}-\hat{\mu})^{2}}{2\hat{\sigma}^{2}}\right)}.\label{eq:Amplitude_Roonizi}
\end{equation}

\vspace{-2.5mm}

\noindent Note that the Roonizi's algorithm has no iterative procedure
to increase the accuracy of fitting long tail GF parameters.

\section*{Motivation}

It is seen that Guo's and Roonizi's\textcolor{blue}{{} }algorithms have
better estimation accuracy than Caruana's algorithm, while their computational
complexity burden is comparable. Moreover, the three algorithms dependently
estimate the GF parameters ($A$, $\mu$ and $\sigma$). This means
that in some applications that require the estimation of only one
parameter, the fitting algorithm may require unnecessary parameters
to be estimated as well. Therefore, there is a need for a new method
which provides better estimation accuracy with an efficient computational
complexity, as well as the capability for a separable parameter estimation.

\vspace{-2mm}

\noindent \begin{flushleft}
\begin{figure}
\begin{centering}
\includegraphics[scale=0.43]{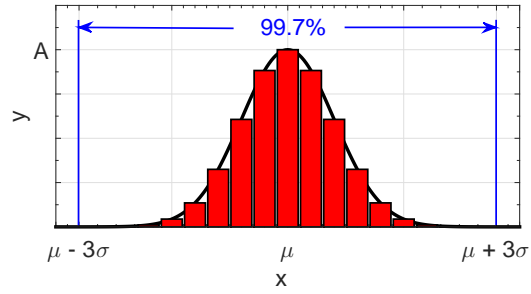}
\par\end{centering}
\centering{}\caption{\label{fig: 1 Gaussian-function.}A Gaussian function.}
\end{figure}
\par\end{flushleft}

\vspace{-2mm}

\section*{Proposed Algorithm}

In this article, we propose a novel computationally efficient (i.e.,
fast), accurate, and separable (FAS) algorithm for a GF that accurately
fits the observed data. The basic idea of the proposed FAS algorithm
is to find a direct formula for the SD (i.e., $\sigma$) parameter
from the noisy observed data, and then, the amplitude $A$ and mean
$\mu$ can be determined using the weighted least-squares method for
only two unknowns.

\vspace{-2mm}

\section*{Derivation of the Standard Deviation Formula}

To derive an approximation formula for the SD, a simple mathematical
trick will be used. For $N$ observations that represent the GF, as
shown in Figure \ref{fig: 1 Gaussian-function.}, the area under the
GF can be divided into thin vertical rectangles with a width of $\Delta x_{n}$,
where $\Delta x_{n}$ is the $n$-th step size of two successive observation
points on the $x$-axis. Therefore, the total area under the GF, $\Lambda$,
is numerically calculated as the summation of the areas of the vertical
rectangles:

\vspace{-2.5mm}

\begin{equation}
\Lambda\approx\sum_{n=1}^{N}\Delta x_{n}\hat{y}_{n}.\label{eq: 22 area under curve numerically}
\end{equation}

\vspace{-2.5mm}

\noindent Note that (\ref{eq: 22 area under curve numerically}) reflects
at least $99.7\%$ of the GF area in case of an available observation
width greater than $\mu\pm3\sigma$. Now, let us calculate the area
under the GF using a different method. From the GF and $Q$-function
properties, the total area under the GF is given as

\vspace{-2.5mm}

\textcolor{black}{
\begin{equation}
\Lambda=\intop_{-\infty}^{\infty}Ae^{-\frac{(x-\mu)^{2}}{2\sigma^{2}}}dx=A\sigma\sqrt{2\pi}.\label{eq:23 integration}
\end{equation}
}

\vspace{-2mm}

\noindent Equating (\ref{eq: 22 area under curve numerically}) and
(\ref{eq:23 integration}), and replacing the amplitude $A$ by the
maximum value of the observed data, $\hat{y}_{\text{max}}$, the estimated
$\sigma$ is obtained as

\vspace{-2.5mm}

\begin{equation}
\hat{\sigma}=\frac{\sum_{n=1}^{N}\Delta x_{n}\hat{y}_{n}}{\sqrt{2\pi}\hat{y}_{\text{max}}}.\label{eq: 25 sigma estimated}
\end{equation}

\vspace{-2.5mm}

\noindent Thus, in certain applications which require the estimate
of the SD of the GF, the FAS algorithm directly outputs this estimate,
without estimating the other two parameters. This is referred to as
the \textit{separable} property of the FAS algorithm.

\vspace{-3mm}

\section*{Error Analysis}

To study the error of (\ref{eq: 25 sigma estimated}), first let us
discuss the systematic error resulting from equating (\ref{eq: 22 area under curve numerically})
and (\ref{eq:23 integration}). This error becomes notable when a
small portion of the GF is sampled, and it is from approximating the
GF curve by rectangles (as in Figure \ref{fig: 1 Gaussian-function.}).
Based on extensive testing of the algorithm with varying parameters,
as discussed further below, the systematic error can be considered
negligible when $W>6$ and the observation samples are dense enough
(e.g., $\frac{N}{W}>10$), where $W$ is the ratio of the SD to the
observation width on the $x$-axis (i.e., the observation width equals
$W\sigma$, or equivalently, it varies from $\mu-\frac{W}{2}\sigma$
to $\mu+\frac{W}{2}\sigma$).

To calculate the relative error in the numerator in (\ref{eq: 25 sigma estimated}),
let the numerator equal $\sqrt{2\pi}A\sigma+\Delta x\sum_{n=1}^{N}w_{n}$,
where $\sqrt{2\pi}A\sigma$ represents the actual area of the GF and
$\Delta x\sum_{n=1}^{N}w_{n}$ is normally distributed with its SD
being $\sqrt{N}\sigma_{w}\Delta x=\sqrt{N}\sigma_{w}\frac{W\sigma}{N}$.
For simplicity of analysis, $\Delta x$ is considered to be fixed
for all observations. The relative error of the numerator, $\alpha_{N}$,
can be written as

\vspace{-2.5mm}

\begin{equation}
\alpha_{N}\approx k_{1}\frac{\sigma_{w}W}{\sqrt{2\pi}A\sqrt{N}}=k_{1}\frac{W}{\text{snr}\sqrt{2\pi N}},\label{eq:Numerator_RE}
\end{equation}

\vspace{-2.5mm}

\noindent where $k_{1}$ is a constant value which can be considered
$2$ for the $95.5\%$ confidence interval, and $\text{snr}=A/\sigma_{w}$
is the signal-to-noise ratio.

For the denominator, let us assume that it equals $\sqrt{2\pi}(A\pm\Delta A)$,
where $\Delta A$ is the maximum of the normally distributed noise
samples with SD of $\sigma_{w}$. The relative error of the denominator
in (\ref{eq: 25 sigma estimated}), $\alpha_{D}$, can be written
as

\vspace{-2.5mm}

\begin{equation}
\alpha_{D}\approx\frac{k_{2}\sigma_{w}}{A}=\frac{k_{2}}{\text{snr}},\label{eq:Denominator_RE}
\end{equation}

\vspace{-2.5mm}

\noindent where $k_{2}$ is a constant whose value can be assumed
to be $3$.\footnote{Based on comprehensive simulations, it is found that $k_{2}=3$ is
the worst-case scenario for the error. Also, the probability of such
a scenario is very low.} Hence, the total relative error in (\ref{eq: 25 sigma estimated}),
$\alpha$, can be approximated using a Taylor series as

\vspace{-2.5mm}

\begin{equation}
\alpha\approx\alpha_{N}+\alpha_{D}=\frac{1}{\text{snr}}\left(k_{1}\frac{W}{\sqrt{2\pi N}}+k_{2}\right).\label{eq:Total_RE}
\end{equation}

\vspace{-2.5mm}

\noindent It is worth noting that if the samples are dense enough
(i.e., large enough $N/W$) and for high snr, a reduced relative error
can be attained.

\section*{Estimates of the Remaining Two Parameters}

To estimate the remaining two parameters $A$ and $\mu$ using $\hat{\sigma}$
estimated from (\ref{eq: 25 sigma estimated}), we can differentiate
the sum of $\delta^{2}$ with respect to $a$ and $b$ and then equate
the results to zero (i.e., using the same steps as in Guo's algorithm).
The resulting linear system of equations becomes

\vspace{-2mm}

\begin{equation}
\left[\begin{array}{cc}
\sum\hat{y}_{n}^{2} & \sum x_{n}\hat{y}_{n}^{2}\\
\sum x_{n}\hat{y}_{n}^{2} & \sum x_{n}^{2}\hat{y}_{n}^{2}
\end{array}\right]\left[\begin{array}{c}
a\\
b
\end{array}\right]=\left[\begin{array}{c}
\sum\hat{y}_{n}^{2}\text{ln}(\hat{y}_{n})-c\sum x_{n}^{2}\hat{y}_{n}^{2}\\
\sum x_{n}\hat{y}_{n}^{2}\text{ln}(\hat{y}_{n})-c\sum x_{n}^{3}\hat{y}_{n}^{2}
\end{array}\right],\label{eq: 26 ibrahim matrix}
\end{equation}

\begin{figure*}
\begin{centering}
\subfloat[${\normalcolor N=50}$]{\begin{centering}
\includegraphics[scale=0.3]{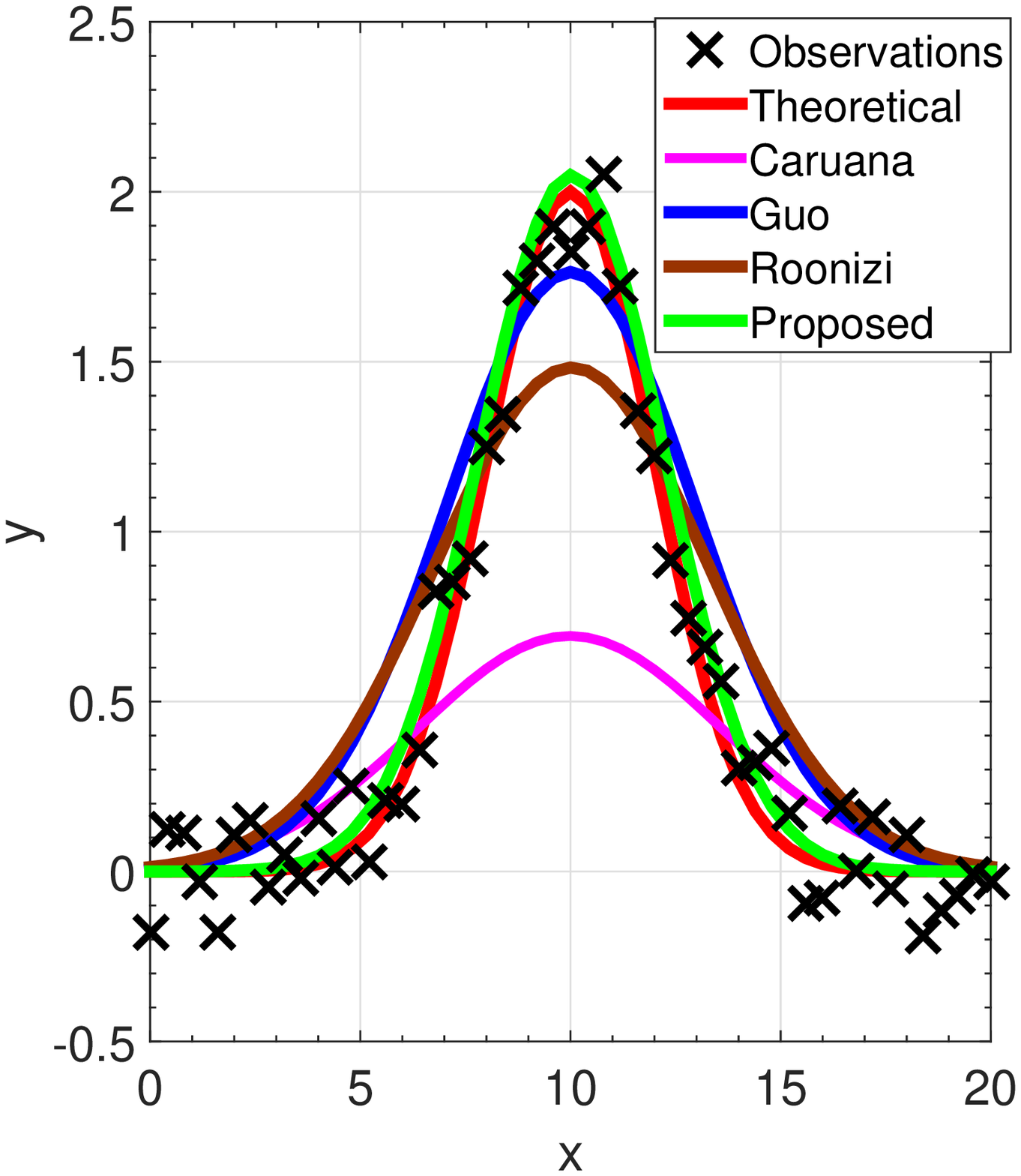}
\par\end{centering}
}\subfloat[$N=40$]{\begin{centering}
\includegraphics[scale=0.3]{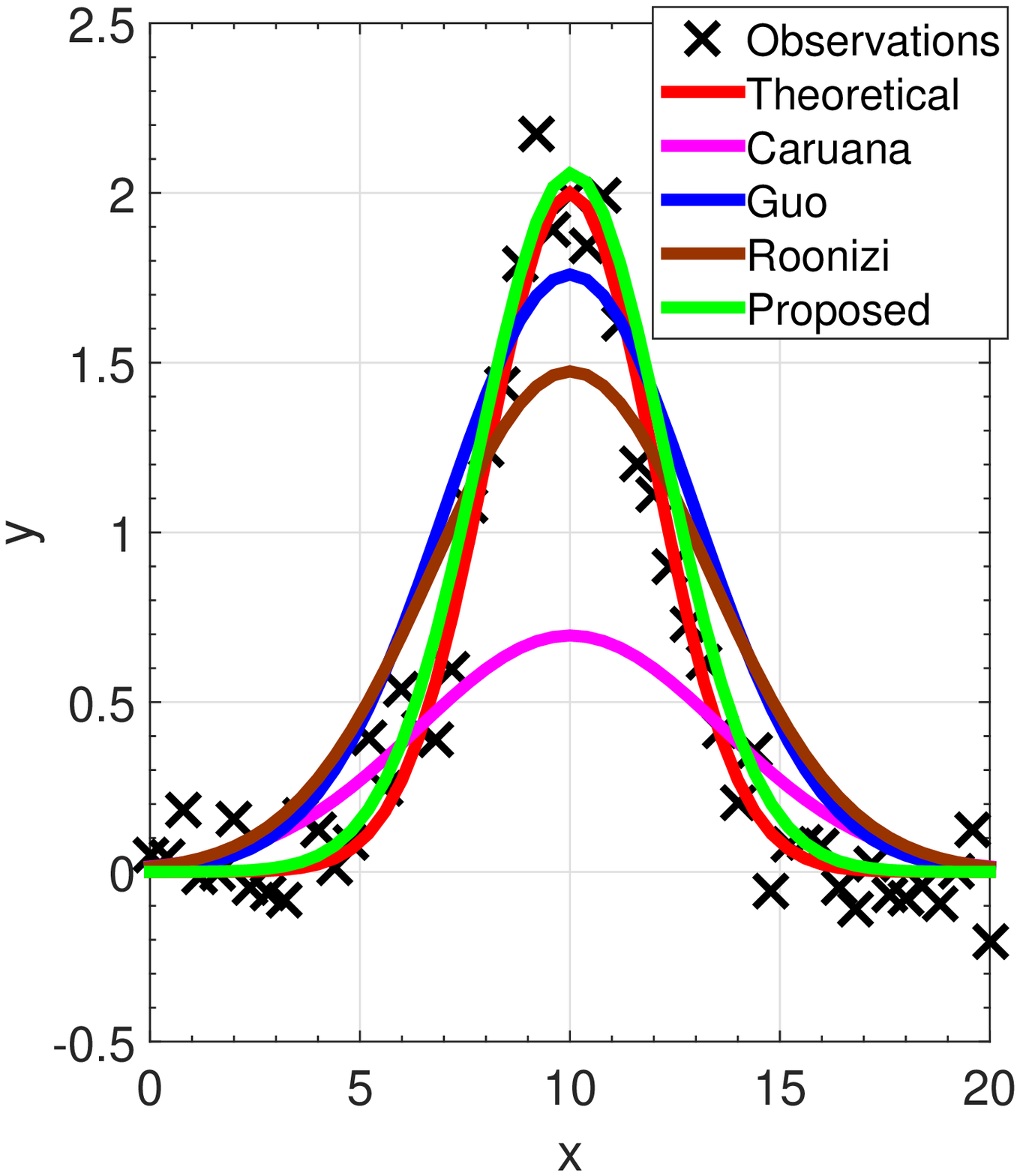}
\par\end{centering}
}\subfloat[$N=30$]{\begin{centering}
\includegraphics[scale=0.3]{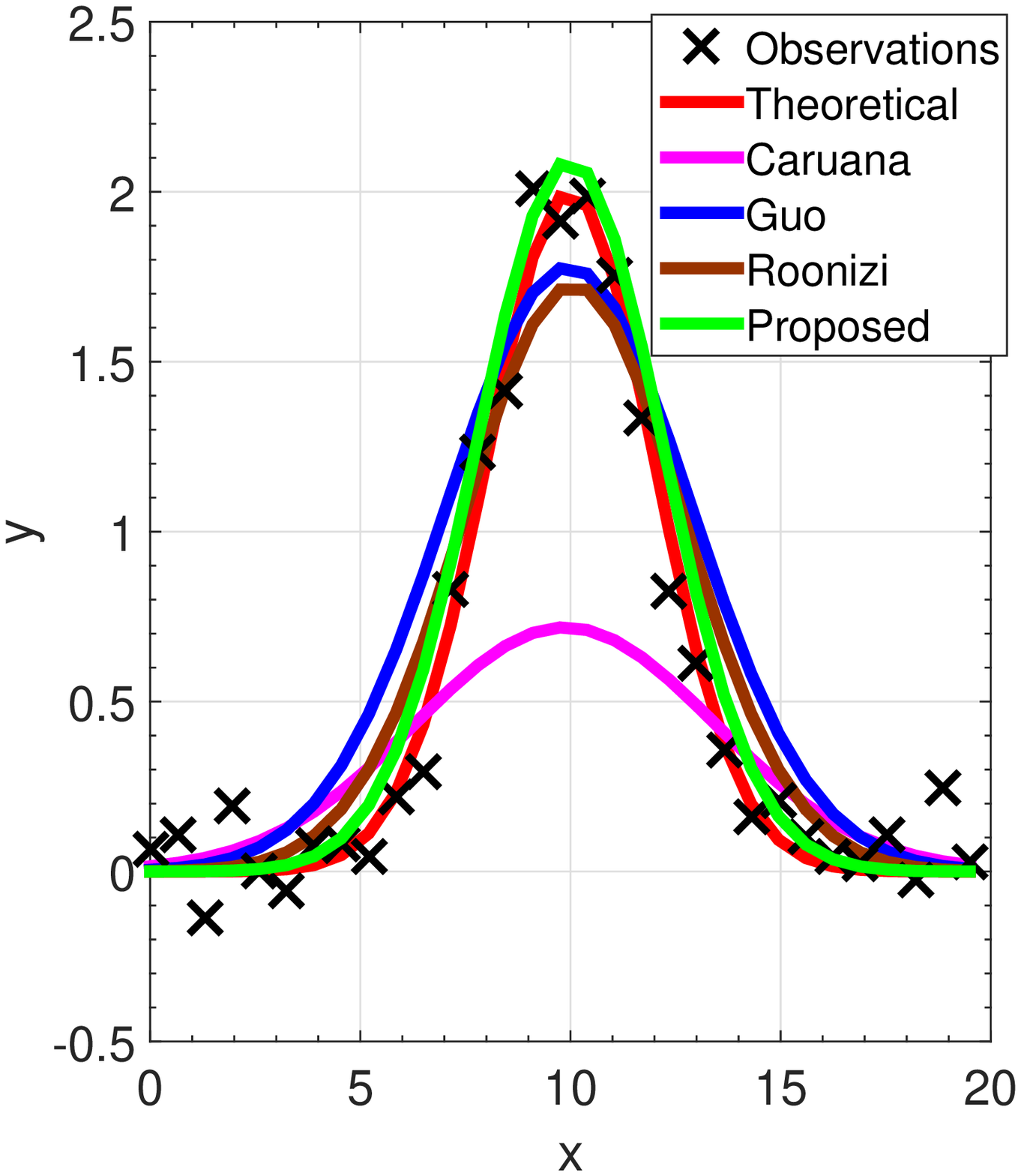}
\par\end{centering}
}
\par\end{centering}
\caption{\label{fig: 2 comparison guassian func}Results of different algorithms
for fitting the GF with $A=2$, $\sigma=2$ and $\mu=10$ in the presence
of observation noise with $\sigma_{w}=0.1$ (i.e., snr = 10).}
\end{figure*}

\vspace{-2mm}

\noindent where $c=-1/\left(2\hat{\sigma}^{2}\right)$ and $\hat{\sigma}$
is the estimated SD, which is calculated from (\ref{eq: 25 sigma estimated}).
Therefore, the values of $a$ and $b$ are obtained by solving the
$2\times2$ linear system in (\ref{eq: 26 ibrahim matrix}); then,
the original parameters $A$ and $\mu$ can be calculated from (\ref{eq:8 mu =000026 sigma =000026 A}).

Figure \ref{fig: 2 comparison guassian func}  shows the superiority
of the proposed FAS algorithm over the traditional algorithms in the
presence of a noise with SD $\sigma_{w}=0.1$ for different values
of $N$; the proposed algorithm provides the best fit to the observed
data points compared to the other fitting algorithms for all values
of $N$. It is seen from Figure \ref{fig: 2 comparison guassian func}
that $\hat{y}_{\text{max}}$ is obviously different from the actual
amplitude $A$. However, $\hat{\sigma}$ from (\ref{eq: 25 sigma estimated})
provides reasonable results using $\hat{y}_{\text{max}}$ even if
a small number of observation points are available as in Figure \ref{fig: 2 comparison guassian func}(c).

Since the FAS algorithm provides poorer accuracy in fitting long tail
GF parameters, an iterative procedure is required to improve the fitting
accuracy.

\vspace{-2mm}

\section*{FAS Algorithm with Iterative Procedure}

For the long tail GF, we propose an iterative algorithm that improves
the fitting accuracy of the FAS algorithm. The recursive version of
(\ref{eq: 26 ibrahim matrix}) is given as

\vspace{-2mm}

\begin{equation}
\left[\hspace{-1mm}\begin{array}{cc}
\sum\hat{y}_{n,(k-1)}^{2} & \sum x_{n}\hat{y}_{n,(k-1)}^{2}\\
\sum x_{n}\hat{y}_{n,(k-1)}^{2} & \sum x_{n}^{2}\hat{y}_{n,(k-1)}^{2}
\end{array}\hspace{-1mm}\right]\left[\hspace{-1mm}\begin{array}{c}
a_{(k)}\\
b_{(k)}
\end{array}\hspace{-1mm}\right]=\left[\hspace{-1mm}\begin{array}{c}
\sum\hat{y}_{n,(k-1)}^{2}\text{ln}(\hat{y}_{n})-c\sum x_{n}^{2}\hat{y}_{n,(k-1)}^{2}\\
\sum x_{n}\hat{y}_{n,(k-1)}^{2}\text{ln}(\hat{y}_{n})-c\sum x_{n}^{3}\hat{y}_{n,(k-1)}^{2}
\end{array}\hspace{-1mm}\right]\hspace{-1mm},\label{eq: iterative ibrahim matrix}
\end{equation}

\noindent where $\hat{y}_{n,(k)}=\hat{y}_{n}$ for $k=0$ and $\hat{y}_{n,(k)}=e^{a_{(k)}+b_{(k)}x_{n}+cx_{n}^{2}}$
for $k>0$, and $\hat{\sigma}$ is estimated from (\ref{eq: 25 sigma estimated})
only once. This means that (\ref{eq: 25 sigma estimated}) can provide
accurate results in fitting the long tail GF without iteration, while
the other two parameters still need to be estimated through iterations.
However, after a few iterations, $\hat{\sigma}$ can be further improved
by including an updated SD from (\ref{eq: 25 sigma estimated}) in
the iterations, using $A$ obtained by (\ref{eq: iterative ibrahim matrix}).

\begin{figure}
\begin{centering}
\subfloat[$\mu=18$]{\includegraphics[scale=0.365]{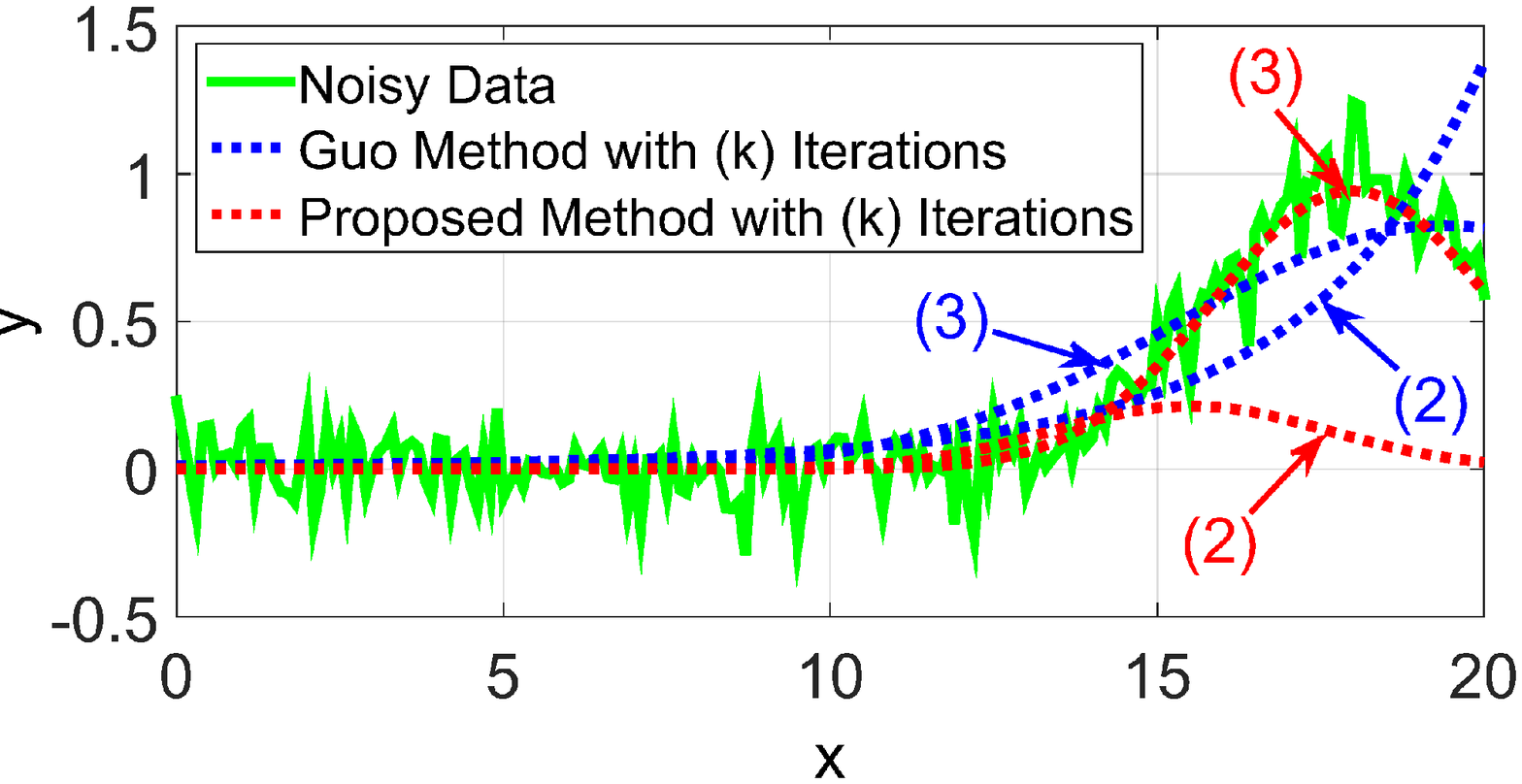}}~~~~~\subfloat[$\mu=19$]{\includegraphics[scale=0.365]{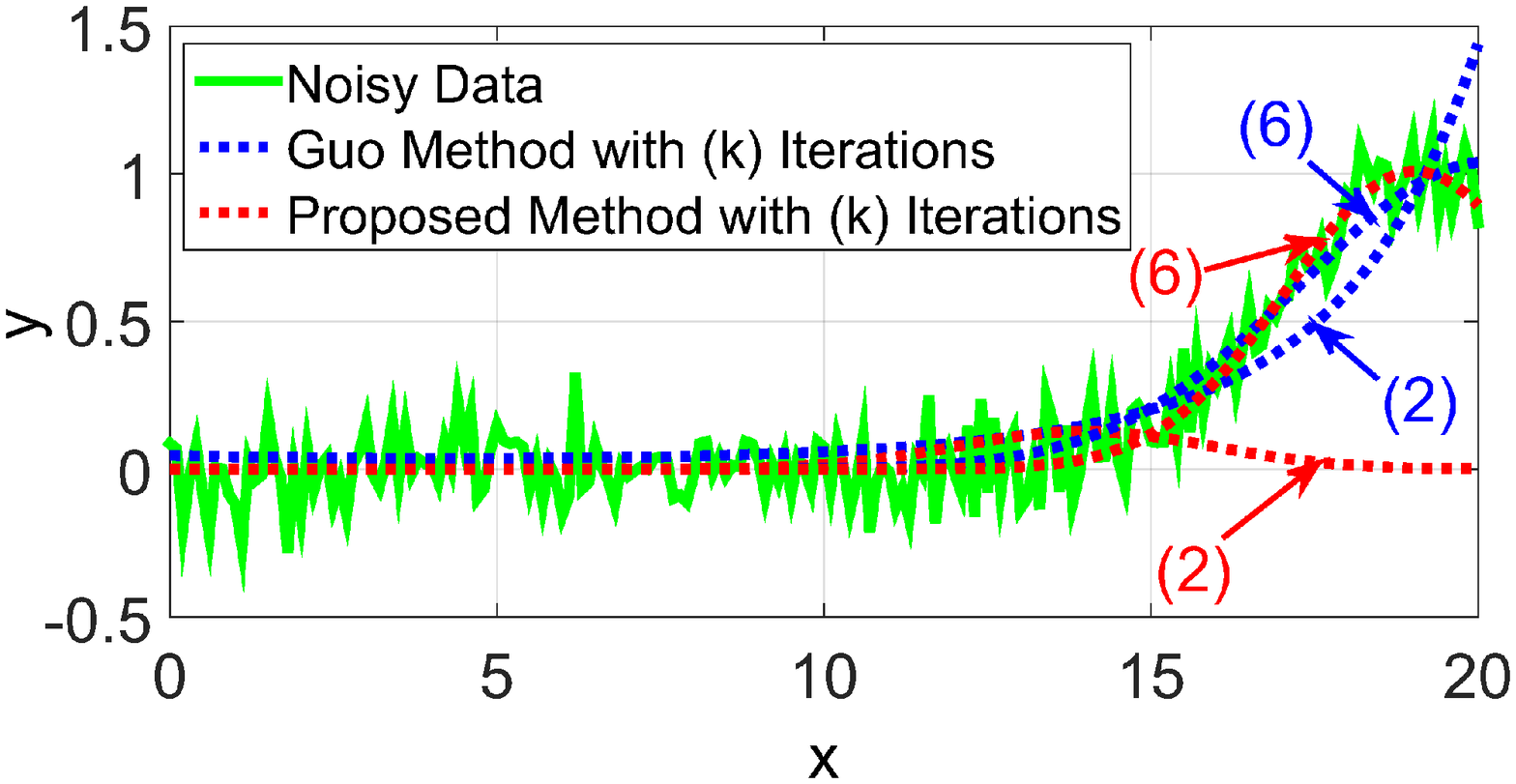}}
\par\end{centering}
\caption{\label{fig: iterative}Results of the proposed FAS iterative algorithm
in comparison with Guo's algorithm for fitting the GF of $N=200$,
$A=1$, $\sigma=2$ and $\sigma_{w}=0.1$ (i.e., snr = 10).}
\end{figure}

Figure \ref{fig: iterative} shows results of the iterative Guo and
proposed FAS algorithms for fitting a long tail GF with $N=200$,
$A=1$, $\sigma=2$ and $\sigma_{w}=0.1$ for $\mu=18$ and $19$,
respectively. As we can see from the figure, the number of iterations
required for the FAS algorithm to fit the long tail GF is lower than
that of Guo's algorithm. For example, in Figure \ref{fig: iterative}(a),
the FAS algorithm needs only 3 iterations to fit the observation;
however, the Guo algorithm provides poor fitting for the same number
of iterations. Note that, from Figure \ref{fig: iterative}(b), the
longer the tail of the GF, the more iterations that are needed (i.e.,
6 iterations are needed instead of 3 to provide a good fitting to
the longer tail GF). It is worthy of noting that in the presence of
large noise and having a small portion of the GF, the iterative procedure
of the proposed algorithm can nicely fit the GF only after a few iterations.

\vspace{-2mm}

\section*{Accuracy Comparison}

\begin{figure}
\begin{centering}
\subfloat[$W=12$ and $N=30$.]{\begin{centering}
\includegraphics[scale=0.33]{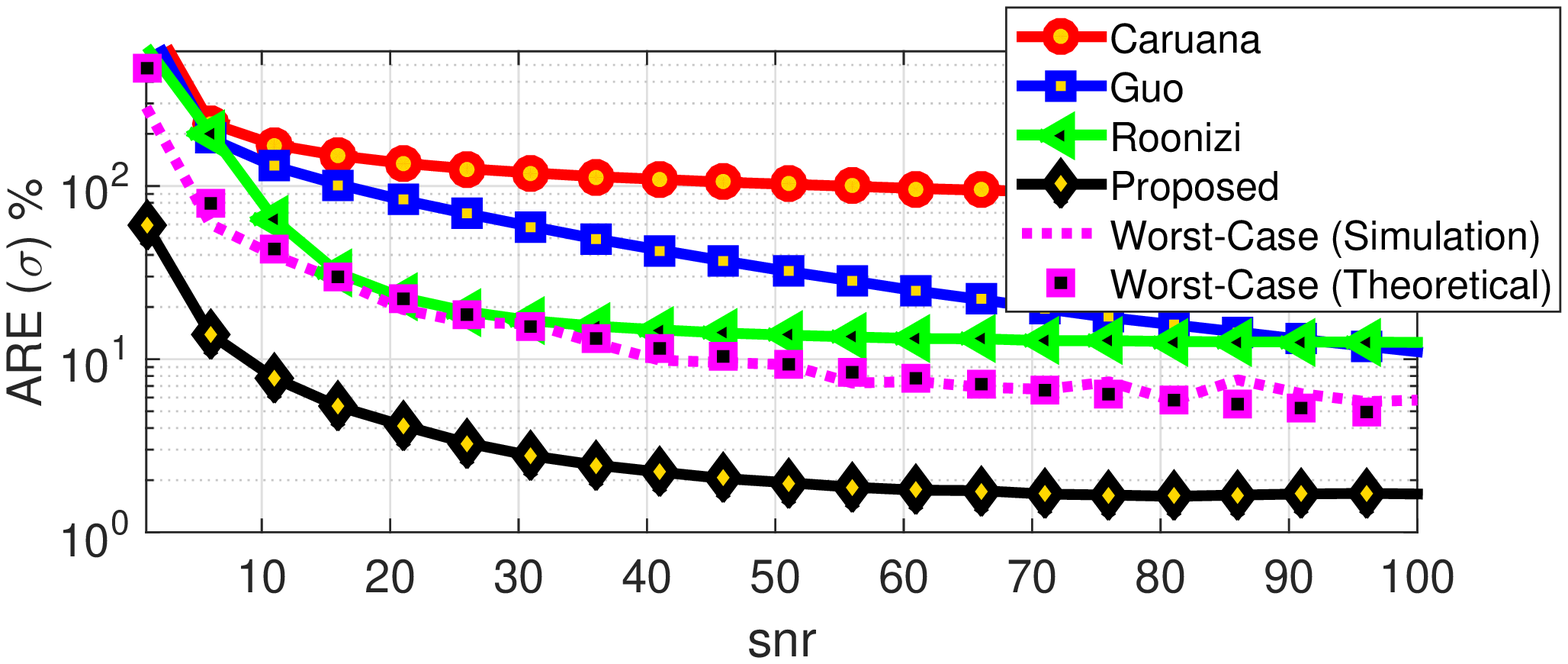}
\par\end{centering}

}\subfloat[$W=12$ and $N=200$.]{\begin{centering}
\includegraphics[scale=0.33]{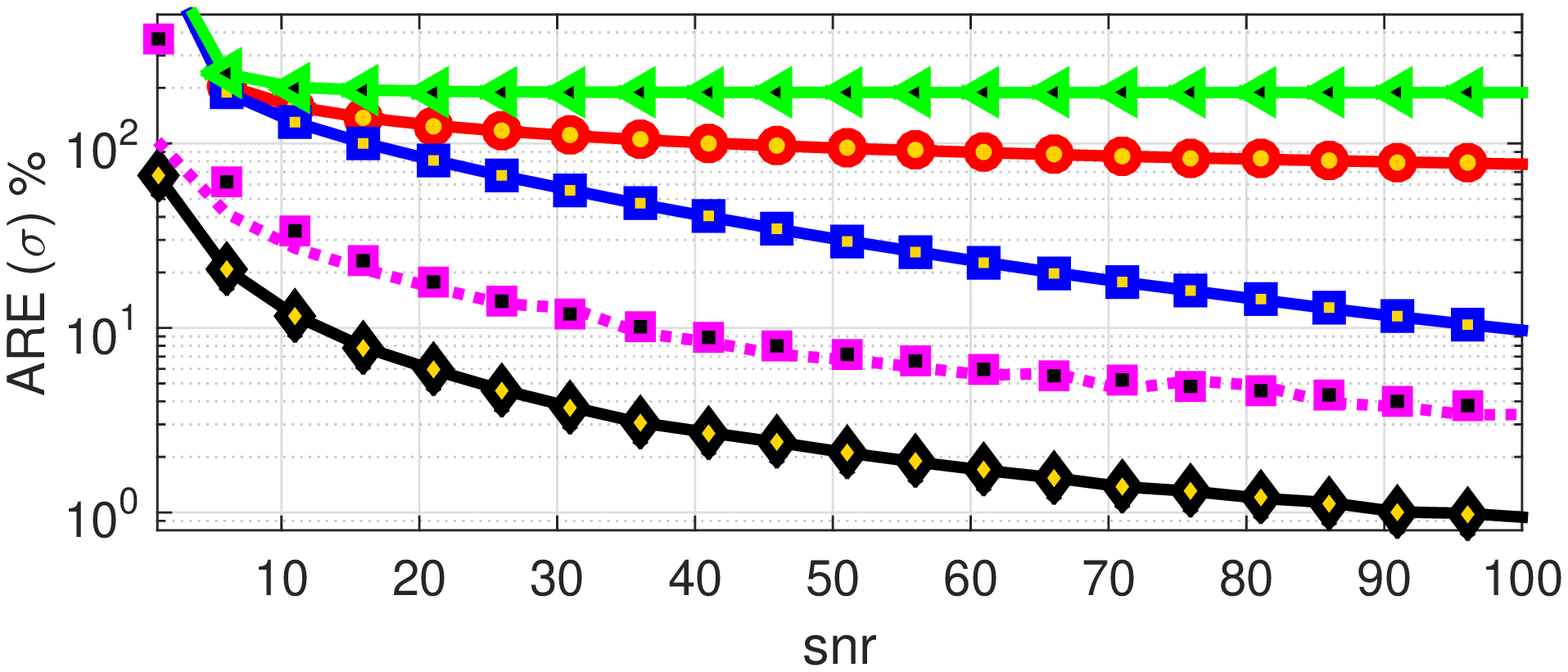}
\par\end{centering}
}
\par\end{centering}
\begin{centering}
\subfloat[snr = 25 and $N=30$.]{\begin{centering}
\includegraphics[scale=0.33]{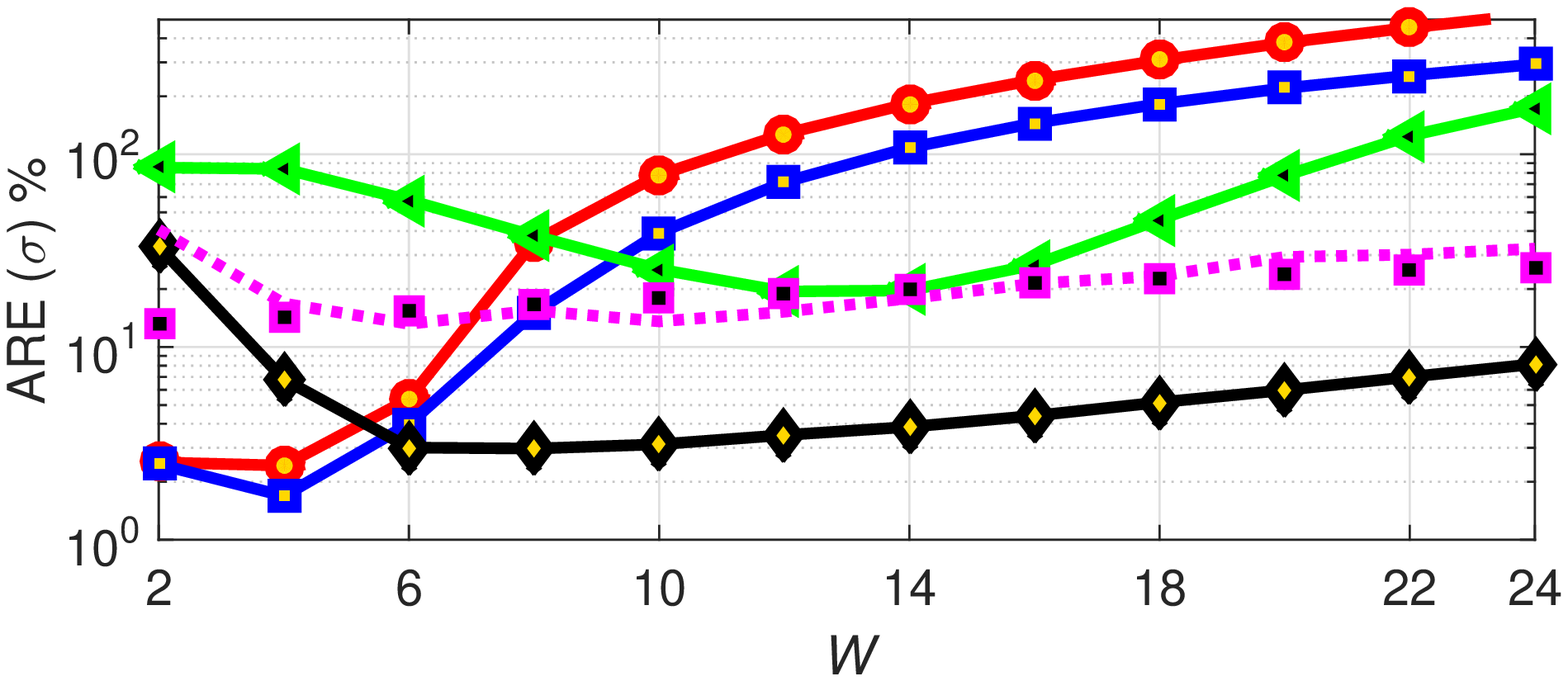}
\par\end{centering}
}\subfloat[snr = 25 and $N=200$.]{\begin{centering}
\includegraphics[scale=0.33]{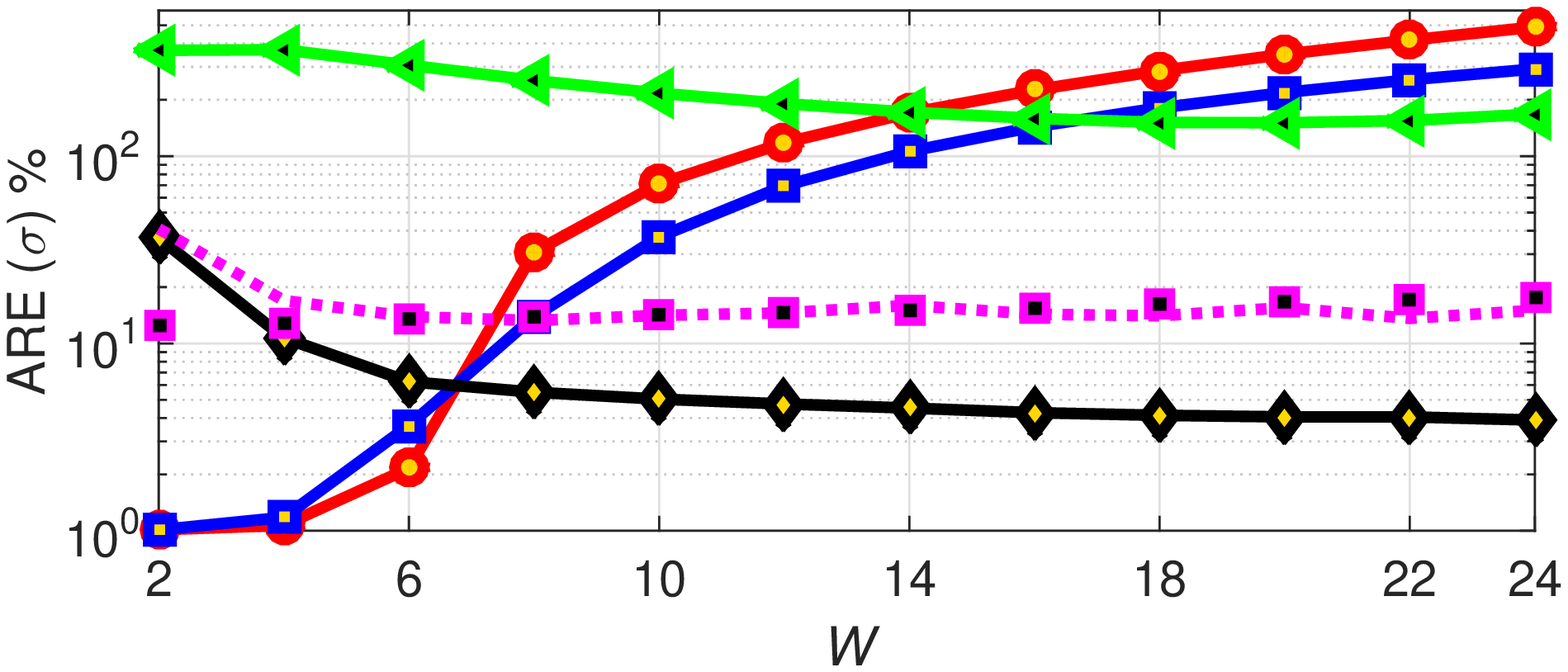}
\par\end{centering}
}
\par\end{centering}
\begin{centering}
\subfloat[$W=12$ and snr = 25.]{\begin{centering}
\includegraphics[scale=0.33]{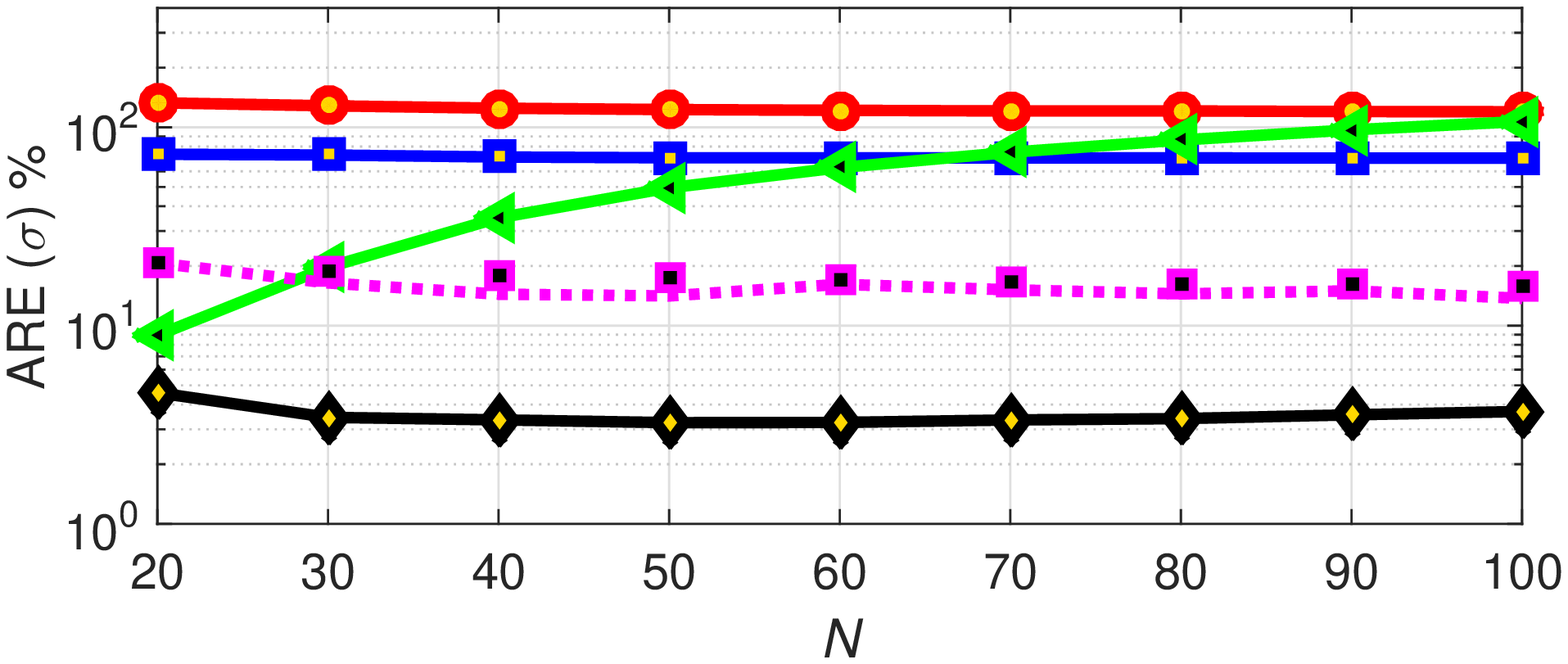}
\par\end{centering}
}\subfloat[$W=12$ and snr = 25.]{\begin{centering}
\includegraphics[scale=0.33]{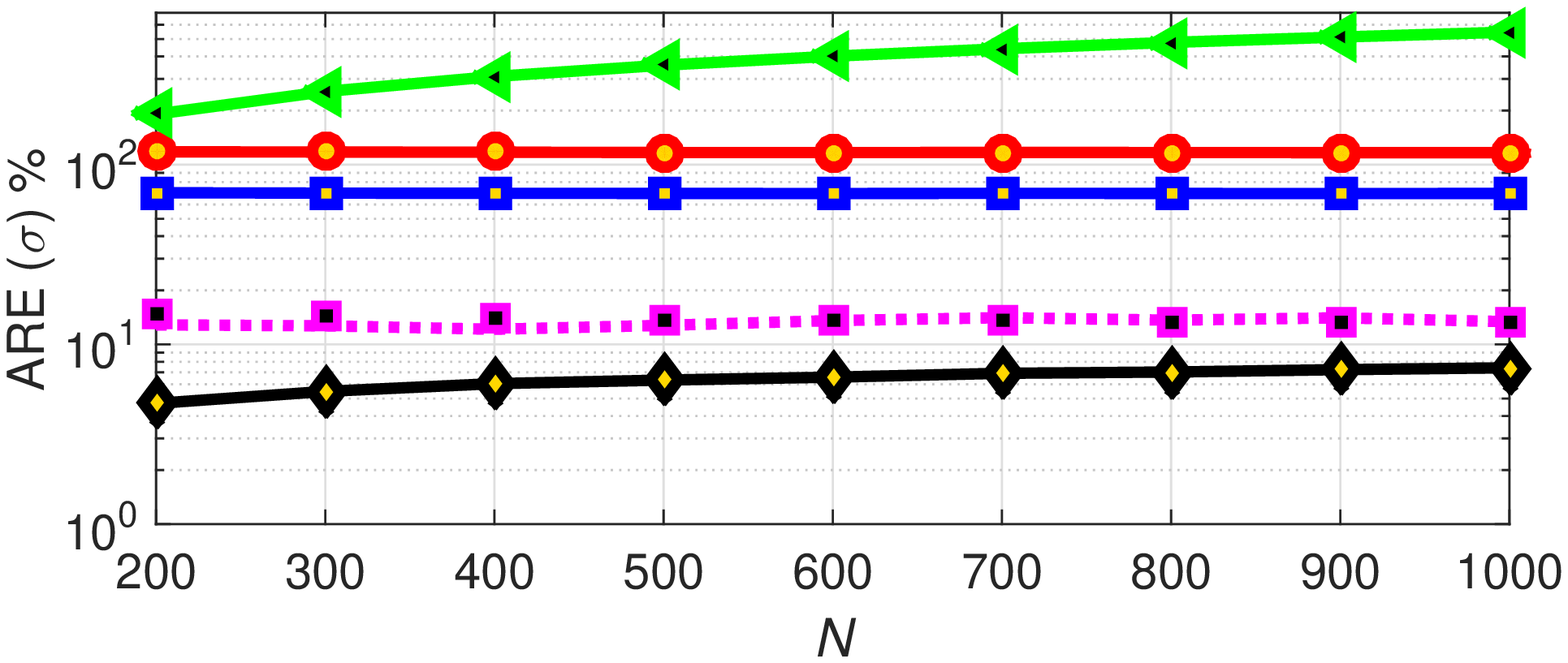}
\par\end{centering}
}
\par\end{centering}
\centering{}\caption{\label{fig:5 variance variations}ARE\% of $\sigma$ estimated from
different fitting algorithms.}
\end{figure}

In this section, Monte Carlo simulation results for at least $10^{4}$
simulated trials are considered to compare the average absolute relative
error (ARE) of the fitting accuracy for the SD estimated using (\ref{eq: 25 sigma estimated})
and by the traditional algorithms. The ARE percentile of the SD is
given as $\text{ARE\%}(\sigma)=\frac{\left|\hat{\sigma}-\sigma\right|}{\sigma}\times100\%$,
where $\left|\cdotp\right|$ denotes the absolute value and $\sigma$
is the true SD. The GF parameters used for this simulation are $A=1$,
$\mu=10$ and $\sigma=2$. As demonstrated by the total relative error
estimated in (\ref{eq:Total_RE}), three parameters can be used for
assessing the accuracy of estimation (i.e., snr, $W$ and $N$).

For the evaluation of the estimation accuracy, we calculate the average
$\text{ARE\%}(\sigma)$, where one of the three parameters is varying
while the other two parameters are kept fixed. Figure \ref{fig:5 variance variations}
shows such results, where the SD is estimated using the proposed FAS
algorithm in comparison with the three previously presented traditional
algorithms. In Figures \ref{fig:5 variance variations}(a) and (b),
$W=12$ and the snr varies from 1 to 100 for $N=30$ and $N=200$,
respectively. Figures \ref{fig:5 variance variations}(c) and (d)
depict the effect of $W$, which varies from $2$ to $24$, for $N=30$
and $N=200$, respectively, in the case of snr = 25 (i.e., $\sigma_{w}=0.04$).
Figures \ref{fig:5 variance variations}(e) and (f) show the effect
of $N$, which varies from $20$ to $100$ and from 200 to 1000, respectively,
with $W=12$ and snr = 25. It is obvious from these figures that the
SD estimated from (\ref{eq: 25 sigma estimated}) has the lowest ARE\%
in all cases, except for $W<6$ when Guo's algorithm is the best.
This is called the \textit{accurate} property of the FAS algorithm.
In many practical applications, an adequate portion of the GF (i.e.,
$W\geq6$) is sampled with more than 200 observation points (i.e.,
$N\geq200$). It is worth noting that Roonizi\textquoteright s algorithm
is more general than the rest of the techniques since it can also
fit a Gaussian riding on a polynomial background. This might explain
its poorer performance in comparison to the other algorithms that
fit a sole GF as described by (\ref{eq:1 gauss form}).

The plots in Figure \ref{fig:5 variance variations} also depict the
worst case ARE\% of the proposed algorithm. The simulated worst-case
ARE\% represents the maximum ARE\% that occurs during the $10^{4}$
simulated trials, which is compared to (\ref{eq:Total_RE}) with $k_{1}=2$
and $k_{2}=3$ to show the accuracy of our derived error estimated
in (\ref{eq:Total_RE}). Note that the probability of such a worst-case
error is very low. Notably, the worst-case theoretical and simulated
ARE\% match, except when $W<6$ due to the considerable systematic
error. It is worth mentioning that the superiority of the proposed
algorithm versus the traditional ones holds for the worse case ARE\%
as well; however, for the clarity of the plots in Figure \ref{fig:5 variance variations},
curves corresponding to the latter algorithms were not included. As
shown in Fig. \ref{fig:5 variance variations}(f), after a particular
value of $N$, the error of the denominator in (\ref{eq: 25 sigma estimated})
becomes dominant. As $N$ increases, there will be many samples around
the peak of the GF, and the ARE\% of the proposed algorithm slightly
increases when $N$ increases, finally approaching the worst case
scenario.

\vspace{-2mm}

\section*{Complexity Comparison}

We address the computational complexity comparison between the Guo,
Roonizi and proposed FAS algorithms, in terms of the number of additions
and multiplications required to complete the fitting procedure. We
assume that subtraction and division operations are equivalent in
complexity to addition and multiplication operations, respectively.
It should be noted that solving an $n\times n$ linear system of equations
using Gauss elimination requires $\left(2n^{3}+3n^{2}-5n\right)/6$
additions and $\left(n^{3}+3n^{2}-n\right)/3$ multiplications \cite{key-8}.
Therefore, the total number of additions ($\text{Add}$) and multiplications
($\text{Mul}$) for the Guo, Roonizi and FAS algorithms are given
as:

\vspace{-2mm}

\begin{singlespace}
\begin{equation}
\text{Ad}\text{d}^{(\text{Guo})}=N(A_{\text{ln}}+8)+3,\,\,\,\,\,\text{Mu}\text{l}^{(\text{Guo})}=N(M_{\text{ln}}+11)+17,\label{eq:17 add =000026 mul guo}
\end{equation}

\end{singlespace}

\vspace{-9mm}

\begin{equation}
\text{Ad}\text{d}^{(\text{Roonizi})}=N^{2}+8N+NA_{\text{exp}}-5,\,\,\,\,\,\text{Mu}\text{l}^{(\text{Roonizi})}=0.5N^{2}+9.5N+NM_{\text{exp}}+9,\label{eq: add =000026 mul Roonizi}
\end{equation}

\vspace{-5mm}

\begin{singlespace}
\textcolor{black}{
\begin{equation}
{\normalcolor {\color{blue}{\normalcolor \text{Ad}\text{d}^{(\text{FAS})}=N(A_{\text{ln}}+8)-3,\,\,\,\,\,\text{Mu}\text{l}^{(\text{FAS})}=N(M_{\text{ln}}+10)+12,}}}\label{eq:27 add =000026 mul ibrahim}
\end{equation}
}
\end{singlespace}

\vspace{-1mm}

\noindent where $A_{\text{ln}}$ and $M_{\text{ln}}$ represent the
number of additions and multiplications required to calculate the
natural logarithm, respectively, while $A_{\text{exp}}$ and $M_{\text{exp}}$
represent the number of additions and multiplications required to
calculate the natural exponential in (\ref{eq:Amplitude_Roonizi}),
respectively. Note that the term of $N^{2}$ in (\ref{eq: add =000026 mul Roonizi})
comes from the calculation of $\phi_{1}(x)$ in (\ref{eq: Phi_Roonizi}),
which requires an accumulated numerical integration of $(u\,y(u))$
from the first observation point to the current value of $x$ for
all $N$ observations.

It can be seen from (\ref{eq:17 add =000026 mul guo})-(\ref{eq:27 add =000026 mul ibrahim})
that the proposed algorithm requires fewer additions and multiplications
when compared with Guo's and Roonizi's algorithms. Assuming $A_{\text{ln}}=A_{\text{exp}}$
and $M_{\text{ln}}=M_{\text{exp}}$, the proposed algorithm saves
six additions and $\mathcal{O}\left(N\right)$ multiplications over
the Guo's algorithm, while it saves $\mathcal{O}(N^{2})$ additions
and multiplications over the Roonizi's algorithm. This is referred
to as the \textit{fast} property of the proposed FAS algorithm.

\vspace{-2mm}

\section*{Conclusion}

This article has proposed a simple approximation expression for the
SD of a GF to fit a set of noisy observed data points. This expression
results from a simple mathematical trick, which is based on the equality
between the area under the GF calculated numerically and based on
the $Q$-function properties. Then, the amplitude and mean of the
GF can be calculated using the weighted least-squares method. Through
comprehensive simulations and mathematical analysis, it has been shown
that the proposed algorithm is not only faster than the Guo and Roonizi
algorithms, but also provides better estimation accuracy when an adequate
interval of the GF is sampled. Additionally, an iterative procedure
is proposed, which is suitable to fit the GF when the observed data
points are contaminated with substantial noise as in the long tail
GF case. It has been shown by extensive computer simulations that
the proposed iterative algorithm fits the GF faster than the iterative
Guo algorithm. The proposed algorithm would be useful for several
applications such as Airy disk approximation, laser transmission welding,
fluorescence dispersion, and many digital signal processing applications.

\vspace{-2mm}

\section*{Acknowledgments}

The authors thank the Editor, Professor Roberto Togneri, and the anonymous
reviewers for their valuable comments that have significantly enhanced
the quality of this article. The authors also gratefully acknowledge
Professor Balazs Bank for his assistance and insightful feedback during
the revision of this article. The authors acknowledge the support
of the Natural Sciences and Engineering Research Council of Canada
(NSERC), through its Discovery program. The work of E. Basar is supported
in part by the Turkish Academy of Sciences (TUBA), GEBIP Programme.

\vspace{-2mm}

\section*{Authors}

\textbf{\small{}Ibrahim Al-Nahhal}{\small{} (ioalnahhal@mun.ca) is
a Ph.D. student at Memorial University, Canada. He received the B.Sc.
and M.Sc. degrees in Electronics and Communications Engineering from
Al-Azhar University and Egypt-Japan University for Science and Technology,
Egypt, in 2007 and 2014, respectively. Between 2008 and 2012, he was
an engineer in industry, and a Teaching Assistant at the Faculty of
Engineering, Al-Azhar University in Cairo, Egypt. From 2014 to 2015,
he was a physical layer expert at Nokia, Belgium. He holds three patents.
His research interests are design of low-complexity receivers for
emerging technologies, spatial modulation, multiple-input multiple-output,
and sparse code multiple access.}{\small\par}

\textbf{\small{}Octavia A. Dobre}{\small{} (odobre@mun.ca) is a Professor
and Research Chair at Memorial University, Canada. She was a Visiting
Professor at Massachusetts Institute of Technology, as well as a Royal
Society and a Fulbright Scholar. Her research interests include technologies
for 5G and beyond, as well as optical and underwater communications.
She published over 250 referred papers in these areas. Dr. Dobre serves
as the Editor-in-Chief of the IEEE Communications Letters. She has
been a senior editor and an editor with prestigious journals, as well
as General Chair and Technical Co-Chair of flagship conferences in
her area of expertise. She is a Distinguished Lecturer of the IEEE
Communications Society and a fellow of the Engineering Institute of
Canada.}{\small\par}

\textbf{\small{}Ertugrul Basar}{\small{} (ebasar@ku.edu.tr) received
the B.Sc. degree (Honours) from Istanbul University, Turkey, in 2007,
and the M.S. and Ph.D. degrees from Istanbul Technical University,
Turkey, in 2009 and 2013, respectively. He is currently an Associate
Professor with the Department of Electrical and Electronics Engineering,
Ko\c{c} University, Istanbul, Turkey and the director of Communications
Research and Innovation Laboratory (CoreLab). His primary research
interests include MIMO systems, index modulation, waveform design,
visible light communications, and signal processing for communications.}{\small\par}

\textbf{\small{}Cecilia Moloney}{\small{} (cmoloney@mun.ca) received
the B.Sc. (Honours) degree in mathematics from Memorial University
of Newfoundland, Canada, and the M.A.Sc. and Ph.D. degrees in systems
design engineering from the University of Waterloo, Canada. Since
1990, she has been a faculty member with Memorial University where
she is now a Professor of Electrical and Computer Engineering. During
2004-2009 she held the NSERC/Petro-Canada Chair for Women in Science
and Engineering, Atlantic Region. Her research interests include nonlinear
signal and image processing methods, signal representations, radar
signal processing, and methods for ethics in engineering and engineering
education.}{\small\par}

\textbf{\small{}Salama Ikki}{\small{} (sikki@lakeheadu.ca) is an Associate
Professor in the Department of Electrical Engineering, Lakehead University,
Canada. He received the Ph.D. degree in Electrical Engineering from
Memorial University, Canada, in 2009. From February 2009 to February
2010, he was a Postdoctoral Researcher at the University of Waterloo,
Canada. From February 2010 to December 2012, he was a Research Assistant
with INRS, University of Quebec, Canada. He is the author of 100 journal
and conference papers and has more than 4000 citations and an H-index
of 30. His research interests include cooperative networks, multiple-input-multiple-output,
spatial modulation, and wireless sensor networks.}{\small\par}

\end{document}